\documentclass[twocolumn]{aastex631}
\usepackage{amsmath}
\usepackage{amssymb}
\usepackage{mathrsfs}
\usepackage{xcolor}
\usepackage{hyperref}
\usepackage{physics}
\usepackage{simplewick}
\usepackage{epsfig}
\usepackage{array}
\usepackage{booktabs}
\usepackage{subfigure}

\newcommand{\C}[1]{\textcolor{blue}{#1}}

\begin{document}

\title{Simulations on the collision between debris stream and outer dusty torus: a possible channel for forming fast-rise and long-delayed radio outburst in tidal disruption events}

\author[0009-0003-0516-5074]{Xiangli Lei}
\affiliation{Department of Astronomy, School of Physics, Huazhong University of Science and Technology, Luoyu Road 1037, Wuhan, China}

\author[0000-0003-4773-4987]{Qingwen Wu$^*$}
\affiliation{Department of Astronomy, School of Physics, Huazhong University of Science and Technology, Luoyu Road 1037, Wuhan, China}

\author[0000-0003-3556-6568]{Hui Li}
\affiliation{Theoretical Division, Los Alamos National Laboratory, Los Alamos, NM 87545, USA}

\author[0000-0002-7329-9344]{Ya-Ping Li}
\affiliation{Shanghai Astronomical Observatory, Chinese Academy of Sciences, Shanghai 200030, People’s Republic of China}

\author[0000-0003-3440-1526]{Wei-Hua Lei}
\affiliation{Department of Astronomy, School of Physics, Huazhong University of Science and Technology, Luoyu Road 1037, Wuhan, China}

\author[0000-0001-7350-8380]{Xiao Fan}
\affiliation{Department of Astronomy, School of Physics, Huazhong University of Science and Technology, Luoyu Road 1037, Wuhan, China}

\author[0000-0002-2581-8154]{Jiancheng Wu}
\affiliation{Department of Astronomy, School of Physics, Huazhong University of Science and Technology, Luoyu Road 1037, Wuhan, China}

\author[0000-0001-5019-4729]{Mengye Wang}
\affiliation{Department of Astronomy, School of Physics, Huazhong University of Science and Technology, Luoyu Road 1037, Wuhan, China}

\author[0009-0002-3470-5052]{Weibo Yang}
\affiliation{Department of Astronomy, School of Physics, Huazhong University of Science and Technology, Luoyu Road 1037, Wuhan, China}

\begin{abstract}
The geometrically thick dusty torus structure is believed to exist in the nuclear region of galaxies (especially in active galactic nuclei, AGNs). The debris stream from a tidal disruption event (TDE) will possibly collide with the dusty torus and produce a transient flare. We perform three-dimensional hydrodynamic simulations to model the dynamical evolution of the interaction between unbound debris and dusty torus. During the continuous interaction, the shocked material will be spilled out from the interaction region and form an outflow. We calculate the temporal evolution of synchrotron emission by assuming that the shock accelerates a fraction of electrons in the outflow into a non-thermal distribution. We find that radio emission from the debris-torus collision generates a steep-rise and slow-decline radio light curve due to the sharp edge and dense gas of dusty torus, where the radio outburst delays the main optical/X-ray outburst by several years or even several tens of years. We apply our model to a TDE that happened in a narrow-line Seyfert I (PS16dtm), where both the radio spectrum and the light curve can be roughly reproduced. Future high-sensitivity, wide-field-of-view radio surveys have the opportunity to detect more such radio flares.

\end{abstract}

\keywords{Active galactic nuclei (16), Tidal disruption (1696)}

\section{Introduction} \label{sec:intro}

Tidal disruption event (TDE) will occur when a star passes close to a supermassive black hole (SMBH) with mass $M_{\rm BH}<sim 10^{8}M_{\odot}$ at the center of a galaxy, where the star will be partially or fully disrupted. After the disruption, approximately half of the stellar debris remains bound to the SMBH, while the other half is ejected and forms an unbound stream on a hyperbolic orbit \citep[e.g.,][]{Rees1988, Evans1989}. Either the accretion of the material by SMBH or the debris collision will possibly produce a strong transient flare in multiwavebands \citetext{\citealp[e.g.,][]{van_Velzen2020, Lu2020}; \citealp[see][for review]{Alexander2020}}. Based on the fallback rate, the SMBH normally accretes at a super Eddington rate in TDEs, which may generate strong winds due to the radiation pressure \citep[e.g.,][]{Strubbe2009,Strubbe2011,Lodato2011}. The blackbody emission from the accretion disk with small BH mass and high accretion rate will radiate at soft X-ray waveband \citep[e.g.,][]{Ulmer1999, Roth2016, Stone2016, Auchettl2017, Dai2018, Pasham2018, Curd2019, Saxton2020}, which may be the mechanism for the TDEs that initially detected in X-ray observations. However, most of the TDEs are observed in optical wavebands, which are explained by the reprocessing of the optically thick envelope or shocks from the debris collisions \citep[e.g.,][]{Guillochon2013, Jiang2016, Piran2015, Roth2016, Lu2020, van_Velzen2020, Liu2021, Steinberg2024}. In last decade, the number of TDE candidates increase rapidly with the advent of wide-field optical transient surveys, such as ASAS-SN \citep[e.g.,][]{Holoien2014, Holoien2016a, Holoien2016b, Holoien2019a, Hinkle2021}, ZTF \citep[e.g.,][]{van_Velzen2019, van_Velzen2021a, Hammerstein2023}, SDSS \citep[e.g.,][]{Abazajian2009, van_Velzen2011}, iPTF \citep[e.g.,][]{Hung2017, Blagorodnova2017, Blagorodnova2019} and Pan-STARRS \citep[e.g.,][]{Gezari2012, Chornock2013, Holoien2019b}. These surveys have brought TDE studies into the era of demographic and population studies \citep[e.g.,][]{Sazonov2021, van_Velzen2021a, Hammerstein2023, Yao2023}. One significant discovery from these demographic and population studies is that TDEs are predominantly observed in post-starburst or ``green valley" galaxies, which suggests that the dynamics of stars in nuclear environments undergo significant evolution during galaxy evolution \citep[e.g.,][]{Hammerstein2021, Yao2023, Wang2024}. Some TDEs also occur in AGNs \citep[e.g.,][]{Blanchard2017, Kankare2017, Liu2020}, although they are difficult to be identified due to selection biases against AGN flares and high levels of AGN obscuration. Future multi-waveband surveys can shed further light on the host properties of TDEs \citep[e.g.,][]{Jiang2021b, Masterson2024}.
 
The radio surveys for TDEs are still lacking, but rapid follow-up observations have led to more than twenty radio detections, which roughly correspond to $\sim$ 20--30\% of all TDEs \citep[e.g.,][]{Alexander2020,Goodwin2023}. 
The most famous example is Sw J1644+57, detected in radio $\sim$ 2 days after its initial gamma-ray discovery, with its multi-waveband spectra likely dominated by a relativistic jet \citep[e.g.,][]{Bloom2011, Zauderer2011,Levan2011,Metzger2012}. In the last several years, more candidates with emission possibly dominated by the relativistic jets have been reported, such as Sw J2058+05 \citep[e.g.,][]{Cenko2012}, Sw J1112-82 \citep[e.g.,][]{Brown2015} and AT 2022cmc \citep[e.g.,][]{Andreoni2022}. The origin of the radio emission of other TDEs is still in hot debate, which is normally delayed from the main outburst for several tens of days to several thousand days. \citetext{\citealp[e.g.,][]{Alexander2016,Krolik2016,Pasham2018,Lu2020,Stein2021,Cendes2022,Goodwin2023}; \citealp[see][for review]{Alexander2020}}. The possible explanations include the interaction of surrounding circumnuclear medium (CNM) with the on-axis/off-axis jet \citep[e.g.,][]{Cendes2021,Horesh2021a,Matsumoto2023,Zhou2024,Sfaradi2024,Sato2024}, or the non-relativistic winds \citep[e.g.,][]{Alexander2016,Lu2020,Mou2022,Bu2023,Matsumoto2024,Zhuang2024}, or the unbound debris \citep[e.g.,][]{Krolik2016,Lei2016,Yalinewich2019,Spaulding2022}, and the delayed jet/outflow formation after the state transition \citep[e.g.,][]{Horesh2021a,Sfaradi2022,Cendes2022,Matsumoto2023}. 

The delayed radio emission may provide insight into the nuclear environments of galaxies through the possible interaction of the outflow with the outer environment \citetext{\citealp[see][for review]{Alexander2020,French2020}}. The SMBH activities are strongly coevolved with the nuclear environments throughout the evolution of galaxies, where the accreted gas is provided from the pc-kpc scale, and the SMBH activities also feedback to the host galaxies \citetext{\citealp[see][for review]{Kormendy2013}}. During the active phase of SMBHs, both optically thick accretion disks and geometrically thick torus are observed to coexist in AGNs \citetext{\citealp[see][for review]{Antonucci1993}}. As the accretion rate decreases, the radiation pressure becomes weaker and the cold gas in dusty torus will move in \citep[e.g.,][]{Pier1992,Hatziminaoglou2008,Kawaguchi2011,Kishimoto2011}. The presence of the accretion disk and torus may alter the observational characteristics of TDEs compared to those occurring in normal galaxies. The possible interaction between the bound debris and the accretion disk has been explored in former works, where the TDE phenomena are much different from that happened in vacuum environments \citep[e.g.,][]{Chan2019,MacLeod2020}. It should be noted that the unbound debris will also possibly collide with the outer geometrically thick torus and lead to a special flare. In this work, we perform the hydrodynamics simulations to explore the dynamical evolution of the interaction of the debris stream and the surrounding dusty torus. We also investigate the synchrotron emission from the non-thermal electrons from the shocked material, where both the radio spectrum and the light curve will be presented. The model description is presented in Section \ref{sec:model}, and the model results and application to a TDE candidate are shown in \ref{sec:Result}. Section \ref{sec:conclusion&discussion} are summary and discussion.

\section{Model} \label{sec:model}

\subsection{Dusty Torus and TDE debris} \label{sec:d&t}
To explore the possible interaction between the debris stream ejected out from TDE and the outer dusty torus, we set up the initial distribution of torus and debris respectively. We describe them respectively in more detail as follows.

For the dusty torus, a Keplerian thick torus in hydrostatic equilibrium is adopted as the initial state \citep{Bannikova2012}, where the geometric structure and density distribution in an equilibrium torus model can be derived \citep[][]{Fishbone1976}. In our work, the maximum gas number density inside the equilibrium torus will be set as a free parameter with typical values close to observations. The inner radius of the torus is set from the observational empirical correlation\citep[e.g.,][]{Namekata2016},
\begin{equation} \label{eq:Rin}
    R_{\text{in}} =  1\ \text{pc} \ {\left( \frac{L_{\text{bol}}}{1.3 \times 10^{46} \text{erg/s}} \right)}^{0.5}, \
\end{equation}
In our simulations, we adopt black hole mass $M_{\text{BH}} = 10^6\ M_\odot$, $f_\text{Edd} = L_{\text{bol}}/L_{\text{Edd}}= 0.01$ as fiducial parameters in our simulations, which corresponds to the inner radius of the torus is $R_{\text{in}} = 0.01$ pc.

After the disruption of a star by the tidal force, roughly half of the stellar mass becomes unbound and flies away with a velocity of several thousand km/s \citep[e.g.,][]{Kenyon2008,Guillochon2016}. The evolution of the debris has been investigated with both the smoothed particle hydrodynamics \citep[e.g.,][]{Coughlin2015, Yalinewich2019} and analytical methods \citep[e.g.,][]{Guillochon2016,Coughlin2023}. In this work, we cannot simulate the disruption and the interaction of debris-torus simultaneously. Therefore, we simply adopt the initial distributions of density and velocity for the debris as provided in \citet{Guillochon2016}, and inject the debris at an inner boundary. To build the debris distribution, we consider a fiducial disrupted star with a mass of $M_\text{s} = 1 M_\odot$, the orbital penetration factor $\beta \equiv r_{\rm t}/r_{\rm p}  = 2$ (the ratio of the tidal disruption to pericenter radius) for a complete disruption \citep[e.g.,][]{Guillochon2013}, and the mass of unbound debris is $0.5 M_\odot$. In simulations, we adopt the unbound debris as a straight cylindrical flow with a solid angle of $\Omega_\text{ud} <sim 2 \times 10^{-3}$ sr, which is consistent with theoretical expectations of $\Omega \sim (M_\text{BH}/M_{\text{s}})^{-1/2} \sim 10^{-3}$ sr for a typical mass ratio of $q = M_\text{BH}/M_{\text{s}} \sim 10^6$ \citep{Khokhlov1996,Strubbe2009,Kasen2010}. Considering the interaction of the debris stream with the circular nuclear medium, the debris matter will disperse into low-density wings and the width of these wings is expected to be approximately several times the width of the unbound debris\citep[e.g.,][]{Yalinewich2019}. We simply set the solid angle of low-density wings $\Omega_\text{w} \sim 2 \times 10^{-2}$ sr with a density of one-thousandth of the central debris density. 

\subsection{Numerical Setup} \label{sec:setup}

We employ the publicly available code \texttt{Athena++} \citep{Stone2020} to conduct three-dimensional hydrodynamics simulations of the interaction between the torus and the unbound debris stream with a resolution of $384 \times 512 \times 2048$ cells in radius, polar angle and azimuthal angle. Our simulation domain spans $\theta \in [\pi/30 , \pi/2]$ , $\phi \in [0, 2\pi]$, and $r \in [0.9R_\text{in},8R_\text{in}]$, where the debris is ejected at $0.9R_\text{in}$ that correspond to the Keplerian orbital periods of $t_{\text{K}} \sim 94$ years. We simulate for $ t_{\text{K}}/2\pi \sim 15$ years until the outermost shocked material generated by the interaction approaches the outer boundary at $8R_{\text{in}}$. To avoid the instability caused by the boundaries near the poles, we neglect the region $\theta\in [0, \pi/30]$, which will not affect our results. Outflow boundary conditions are used in the $\theta$ direction and radial outer boundary, and periodic boundary conditions are used in the $\phi$ direction. We inject the debris stream from the inner boundary in the $r$ direction with $\theta_{\text{D}} = 0.4$ (angle between the debris stream and the equatorial plane). The evolution of the density and the velocity are determined based on the model of \citet{Guillochon2016}, where the initial parameters are provided in Section \ref{sec:d&t}. 
We adopt the density distribution of the torus based on the equilibrium torus solutions of \citet{Fishbone1976} and set the initial maximum gas number density inside the torus as $n_{\rm T} = 10^8 \text{cm}^{-3}$ \citep{Uematsu2021}. The background density is set as $n_0 = 1\times 10^3 \text{cm}^{-3}$, which is similar to that determined from X-ray measurements of Bremsstrahlung emission near the Bondi radius of Sgr A* \citep{Guillochon2016,Krolik2016,Yalinewich2019}.

\subsection{Synchrotron Radiation} \label{subsec: SynchrotronRadiation}

\begin{figure} 
  \centering
  \includegraphics[width=1.0\linewidth]{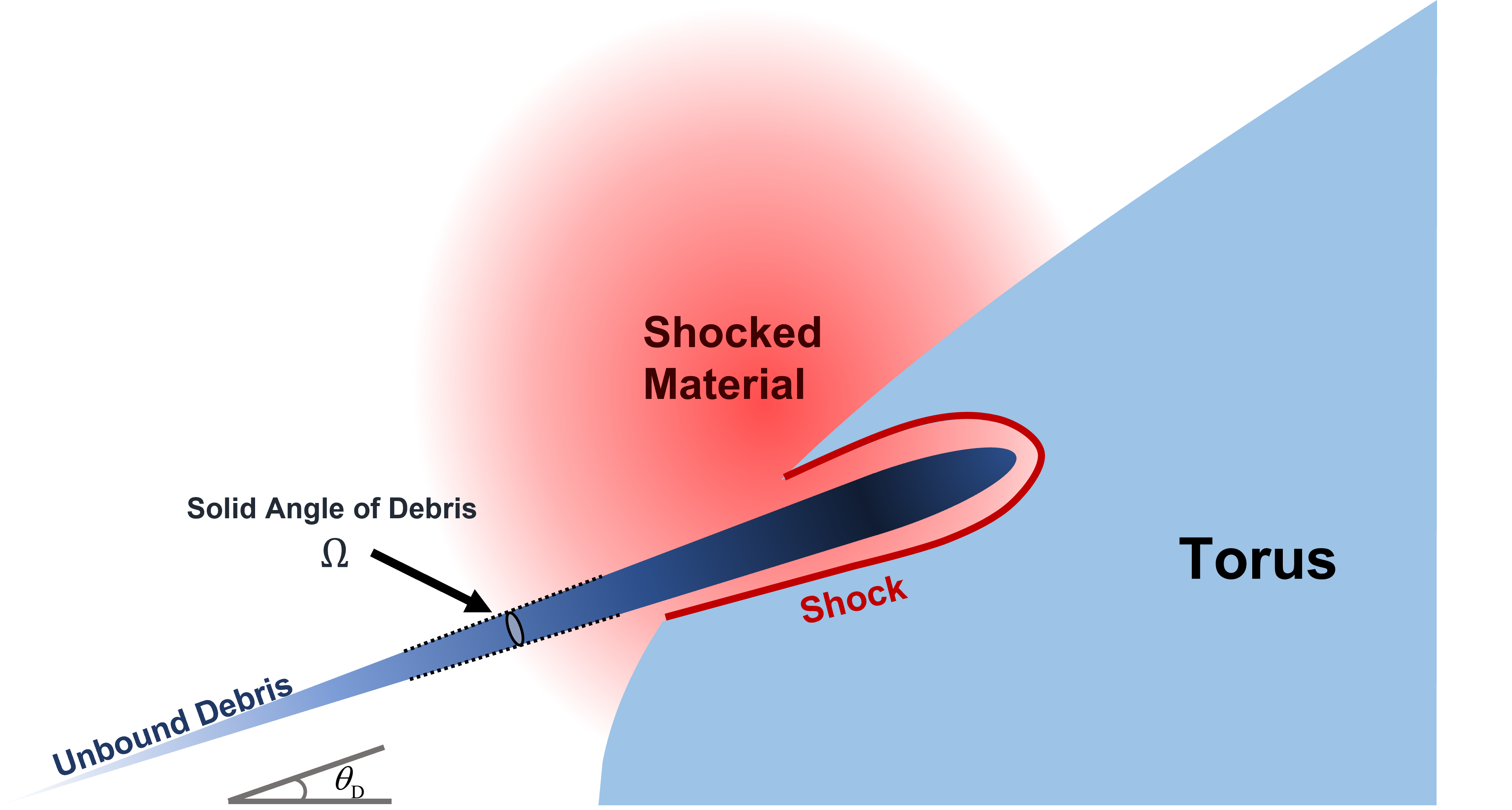}
  \caption{Schematic diagram of the interaction between the unbound debris and the outer dusty torus.}
  \label{fig:cartoon}
\end{figure}

The interaction between the debris stream and the dusty torus will drive shocks. The shocks rapidly propagate into the torus due to the density of the debris stream ($n_{\text{D,max}} \gtrsim 3 \times 10^{10} \text{cm}^{-3}$) is much higher than that of the dusty torus in most cases. The collision will also lead to a fraction of the material being expelled, which will form an outflow (see Figure \ref{fig:cartoon}).  

It is widely believed that shocks can amplify the surrounding magnetic field \citep[e.g.,][]{Reynolds1981}. Following \citet{Sarbadhicary2017}, the magnetic field of the dispersed shocked material can be estimated as
\begin{equation}
    \frac{B^2}{8\pi} = \bar{\epsilon}_\text{B} n_{\text{ds}} m_{\text{p}} v_{\text{ds}}^2,
    \label{eq:B}
\end{equation}
where $n_{\text{ds}}$ represents the number density of dispersed shocked electrons, $v_{\text{ds}}$ is their velocity, $m_{\text{p}}$ is proton mass, $\bar{\epsilon}_\text{B}=\left(v_{\text{ds}}/v_{\text{ds,max}}\right)\epsilon_\text{B}$ is the amplification of the magnetic field by the shocks \citep{Bell2004,Sarbadhicary2017}, $v_{\text{ds,max}} \sim 9 \times 10^8 \text{cm/s}$ is the maximum velocity of dispersed shocked material obtained from the simulation under our typical parameters when debris first hit the torus, and $\epsilon_\text{B}$ is a free parameter describing the strength of magnetic field.

The material undergoing shocks will generate a population of non-thermal electrons, which will follow a power-law distribution $dn_{\text{nt}}(\gamma_{\text{e}}) = A\gamma_{\text{e}}^{-p}d\gamma_{\text{e}}$, where $p=2.1$ is adopted in our model. The minimum electron Lorentz factor, $\gamma_{\rm m}$, depends on the bulk velocity of the shocked material. There is a critical velocity of the dispersed shocked electrons $v_{\text{DN}} = c\sqrt{8m_{\text{e}}/(m_{\text{p}}\bar{\epsilon}_{\text{e}})}$, where $\bar{\epsilon}_{\text{e}} = 4\epsilon_{\text{e}}(p-2)/(p-1)$ is the fraction of shock internal energy that converted into non-thermal electrons energy \citep{Sironi2013}. In our case, the velocity of the dispersed shocked electrons, $v_{\rm ds}$, is normally lower than the bulk velocity of shocked materials, $v_{\text{DN}}$, which is in Deep-Newtonian phase and $\gamma_{\text{m}}$ set to 2 \citep[e.g.,][]{Huang2003}. In the Deep-Newtonian phase, only a fraction $(v/v_{\text{DN}})^2$ of electrons in the shocked material will be accelerated into power-law distribution, and the energy of the nonthermal electrons with $\gamma_{\rm e} > \gamma_{\rm m}$ is  
\begin{equation}
    dE_{\rm nt} = dN_{\text{nt}}m_{\text{e}}c^2\gamma_{\text{m}}\frac{p-1}{p-2} \simeq \epsilon_{\text{e}} dN_{\text{ds}}m_{\text{p}}v_{\text{ds}}^2,
\end{equation}
where $dN_{\text{nt}}$ and $dN_{\text{ds}}$ represent the total number of non-thermal electrons and dispersed shocked electrons in each simulation cell respectively. The coefficient $A$ for electron power-law density distribution is 
\begin{equation}
    A = (p-1)\gamma_{\text{m}}^{p-1}n_{\text{ds}}v_{\text{ds}}^2/v_{\text{DN}}^2.
\end{equation}



For an electron with $\gamma_{\text{m}}$, the typical synchrotron emission frequency is
\begin{equation}
    \nu_{\text{m}} = \frac{eB}{2\pi m_{\text{e}}c} \gamma_{\text{m}}^2.
    \label{eq:nu_m}
\end{equation}
The synchrotron radiation luminosity produced from each simulation cell in the shocked region at this frequency is
\begin{equation}
    dL_{\nu_{\text{m}}} \simeq dN_{\text{nt}} \frac{(4/3)\sigma_Tc\gamma_{\text{m}}^2(B^2/8\pi)}{\nu_{\text{m}}}.
    \label{eq:L_nu_m}
\end{equation}

The synchrotron self-absorption (SSA) frequency $\nu_{\text{a}}$ is normally higher than $\nu_{\text{m}}$ in our case, which is determined by the optical depth of the emission region $\int_l \alpha_\nu \ dl = 1$, where the absorption coefficient $\alpha_\nu \simeq eA(\pi^{3/2}/4)3^{(p+1)/2}B^{-1}\gamma_{\text{m}}^{-p-4}\nu_{\text{m}}^{(p+4)/2}\nu^{-(p+4)/2}$ and the integral is along the line of sight \citep[e.g.,][]{Rybicki1979,Mou2022}. The value of $\nu_{\text{a}}$ is 
\begin{equation}
    \nu_{\text{a}} = \left( \int_l \frac{(p-1)en_{\text{ds}}\pi^{3/2}3^{(p+1)/2}v_{\text{ds}}^2 \nu_{\text{m}}^{(p+4)/2} }{4Bv_{\text{DN}}^2\gamma_{\text{m}}^5} \ dl \right)^{2/(p+4)}. 
    \label{eq:nu_a}
\end{equation}

In this work, we simply assume that the line of sight is perpendicular to the equatorial plane of the accretion disk/torus. The corresponding synchrotron luminosity is $dL_{\nu_{\text{a}}} = dL_{\nu_{\text{m}}} (\nu_{\text{a}}/\nu_{\text{m}})^{(1-p)/2}$. The spectrum of the synchrotron radiation taken into account of SSA is
\begin{equation}
    dL_\nu = 
    \begin{cases}
        dL_{\nu_{\text{a}}} \left( \frac{\nu_{\text{m}}}{\nu_{\text{a}}} \right)^{\frac{1}{2}} \left( \frac{\nu}{\nu_{\text{a}}} \right)^2 & , \quad \nu \leq \nu_{\text{m}} ; \\
        dL_{\nu_{\text{a}}} \left( \frac{\nu}{\nu_{\text{a}}} \right)^{\frac{5}{2}} & , \quad \nu_{\text{m}} \leq \nu \le \nu_{\text{a}} ;  \\
        dL_{\nu_{\text{a}}} \left( \frac{\nu}{\nu_{\text{a}}} \right)^{\frac{1-p}{2}} & , \quad \nu_{\text{a}} \leq \nu . 
    \end{cases}
    \label{eq:L_nu}
\end{equation}

The total synchrotron radiation $L_{\nu,\text{total}}$ for the shocked material is obtained by summing the luminosities from each simulation cell in the shocked region ($L_{\nu,\text{total}} = \sum dL_{\nu}$). 


\section{Results} \label{sec:Result}

\subsection{The dynamic evolution of unbound debris-torus interaction} \label{sec:BasicResult}

\begin{figure*} 
  \centering
  \includegraphics[width=1.0\linewidth]{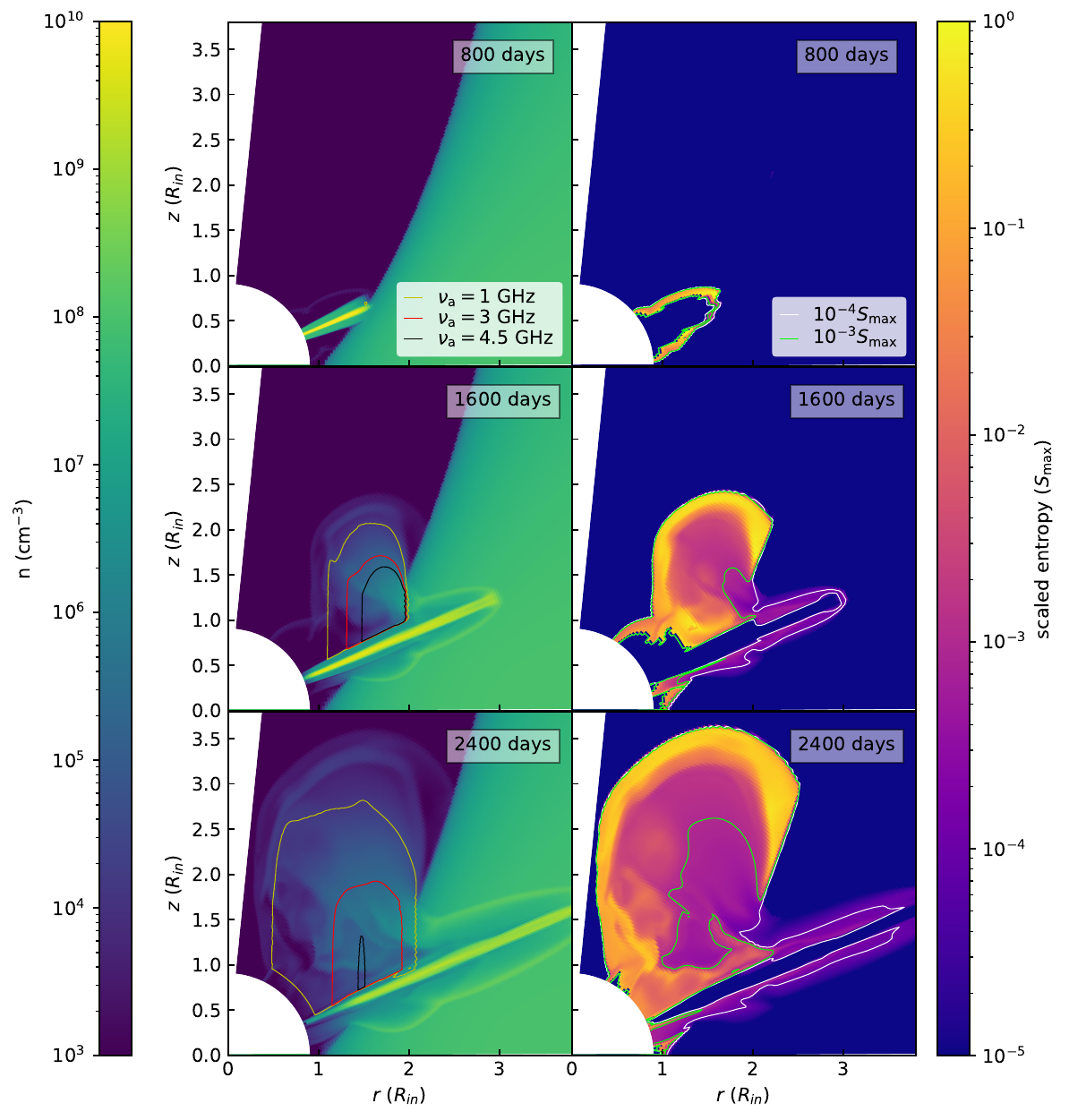}
  \caption{The distributions of the number density and entropy for the materials at the azimuth of the debris stream in our simulation at several typical snapshots. In the left panels, the solid lines representing the optical depth $ \tau = 1 $ at different frequencies: yellow (1 GHz), red(3 GHz), and black(4.5 GHz), where a face-on observer is assumed. In the right panels, the white and green solid lines correspond to $S = 10^{-3} S_{\text{max}}$ and $S = 10^{-4} S_{\text{max}}$, respectively.}
  \label{fig:simulation}
\end{figure*}

 The shock will be formed when the debris streams collide with the outer dusty torus, which will convert the kinetic energy of the debris stream into thermal energy and accelerate a fraction of electrons into non-thermal distribution. In our simulations, we find that the strong shocked material will rebound from the torus and expand outward in the initial stage. As the unbound debris penetrates the torus, a part of the shocked material sweeps through the interior of the torus, where it mixes with the internal torus material. A fraction of this mixed material will flow out from the interacting gap. We present the local number-density distribution for several typical snapshots in Figure \ref{fig:simulation}. These snapshots correspond to different stages of the interaction process: the debris reaching the torus (top-left panel), the outward flow of a fraction of shocked material (middle-left panel), and the expansion of a large amount of shocked material (bottom-left panel).

In simulations, we find that the dispersed shocked material is partly contributed from the direct collision of debris onto the torus, and the other part is contributed from the outward propagation of the shock in the cavity, which was defined as the strong shocked material and the weak shocked material respectively. To better discriminate the strong shocked material and weak shocked material, we present the distribution of the entropy $s = P/\rho^\gamma$ in Figure \ref{fig:simulation}, where $\gamma = 5/3$ is the adiabatic index. To mitigate the impact of shocks generated by the interaction between unbound debris and CNM, we plotted the figures by omitting the low-density regions with $n < 5 \times 10^2 \rm cm^{-3}$. The strong shocked material and weak shocked material also have much different entropy, which can be evidently distinguished from the different colors in the right panels of Figure \ref{fig:simulation}. To better distinguish between the two types of shocked material, we plot two contour lines corresponding to $S = 10^{-3} S_{\text{max}}$ and $S = 10^{-4} S_{\text{max}}$ in right panels of Figure \ref{fig:simulation}, where $S_{\text{max}}$ is the maximum entropy of the shocked material during the entire interaction. It should be noted that the value of $S_{\text{max}}$ is a constant value during each run but different for each parameter sets.


%

\begin{figure} 
  \centering
  \includegraphics[width=1.0\linewidth]{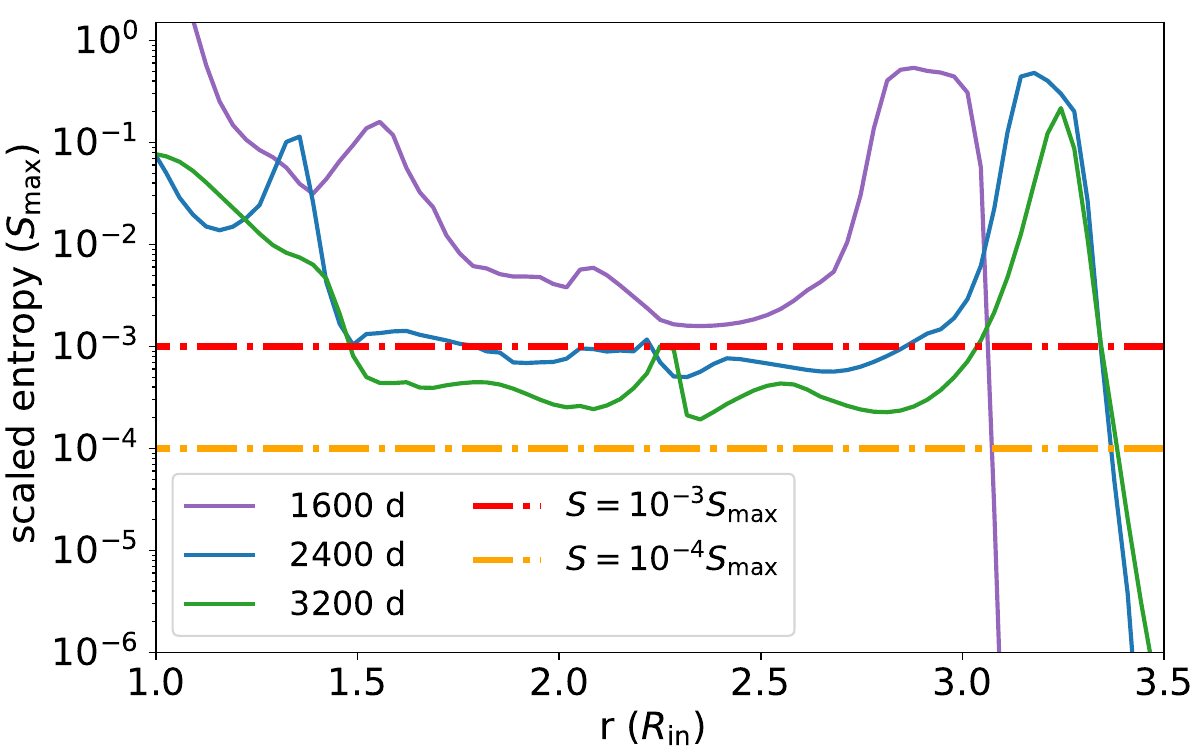}
  \caption{The distributions of the scaled entropy at fixed $\theta = 0.8$, $\phi = 0$ (along the debris) at several typical snapshots, and the red and orange dashed lines represent the $S = 10^{-3} S_{\text{max}}$ and $S = 10^{-4} S_{\text{max}}$, respectively.}
  \label{fig:entropy}
\end{figure}

In the first snapshot of Figure \ref{fig:simulation} ($t = 800$ days, top-right panel), the slowly diffused strong shocks are triggered by the interaction between the debris stream and the background material, where the debris has not yet collided with the dusty torus. In the second snapshot ($t = 1600$ days, middle-right panel), the interaction produces strong shocks and a fraction of shocked material rebounds into interstellar space, where the head of the unbound debris starts to collide with the surface of the torus. After this period, the outward expanding shocked material has higher entropy values, which is more evident in subsequent snapshots. As the unbound debris continues to penetrate into the dusty torus, the lower-density wings are obstructed by the torus and continuously generate strong shocks, which rebound material into interstellar space. Near the point of interaction between the wings and the torus surface, some fraction of the torus material will mix with the shocked material, which will reduce the entropy. The high-density debris will carve out a cavity inside the torus, where shocked material mixes with the unshocked material of the torus and slowly flows outward from the cavity. The outflowing material from the cavity normally has a lower entropy (see bottom-right panels of Figure \ref{fig:simulation}). In Figure \ref{fig:entropy}, we present the entropy distribution along a selected line a fixed $\theta = 0.8$, where the material in the outer shell has high entropy(e.g., $\sim 10^{-3}S_{\text{max}}-10^{-1}S_{\rm max}$) and the material in the inner part normally have lower entropy (e.g., $\sim 10^{-4}S_{\text{max}}-10^{-3}S_{\rm max}$). For simplicity, we define the material with $S\ge 10^{-3} S_{\text{max}}$ and  $10^{-4} S_{\text{max}} \le S \le S_{\text{max}}$ as strong shocked material and weak shocked material respectively. In this work, we neglect the emission from the material with $S\le 10^{-4} S_{\text{max}}$, where the shock should be very weak compared above two cases. It should be noted that the critical entropy to distinguish strong or weak shocked material ($\sim 10^{-3}S_{\text{max}}$) is roughly similar in different simulation sets. It will not affect our main conclusion even if it may be slightly changed, which will be discussed later.


\subsection{Radio light curve} \label{sec:TypicalParameters}


\begin{figure} 
  \centering
  \includegraphics[width=1.0\linewidth]{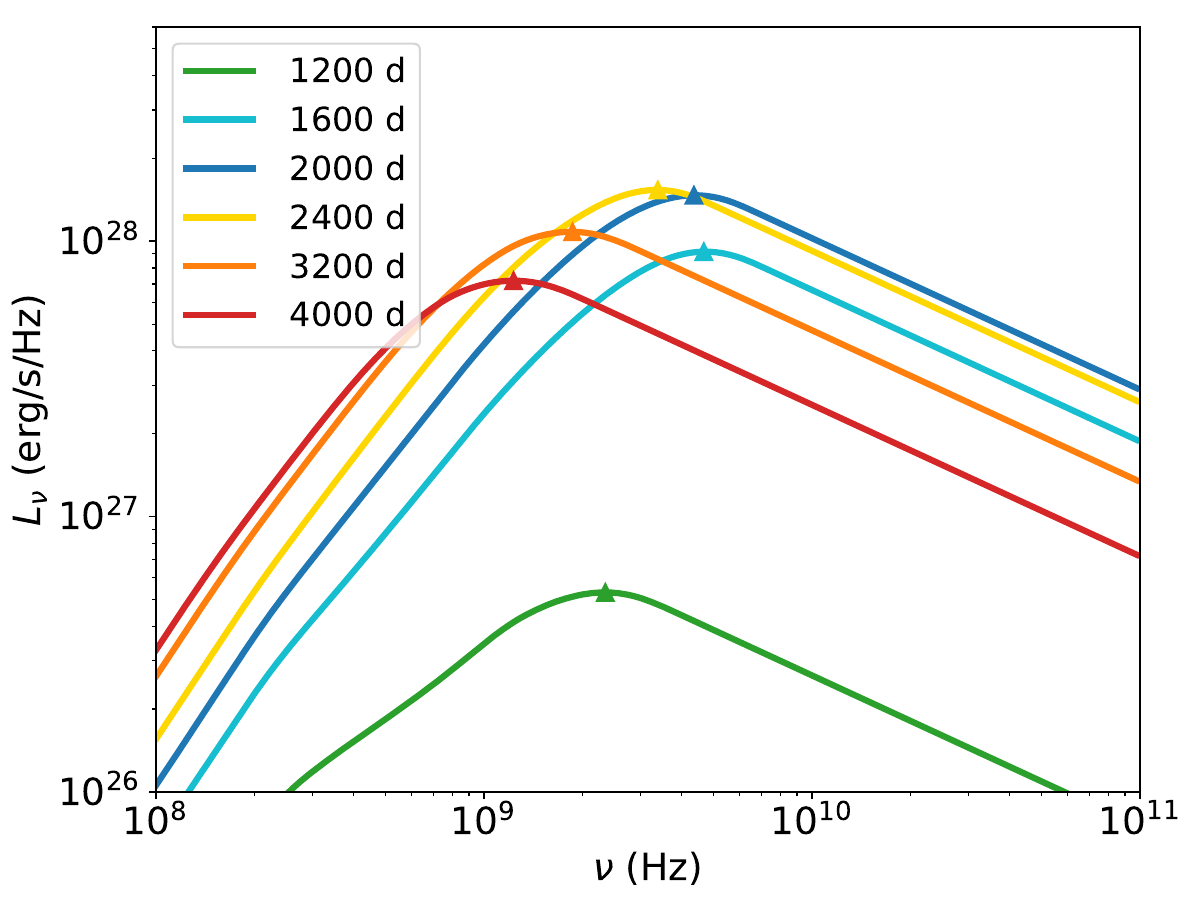}
  \caption{The radio spectrum at different epochs, where triangles correspond to the peak luminosity.}
  \label{fig:spc}
\end{figure}

\begin{figure*} 
  \centering
  \includegraphics[width=1.0\linewidth]{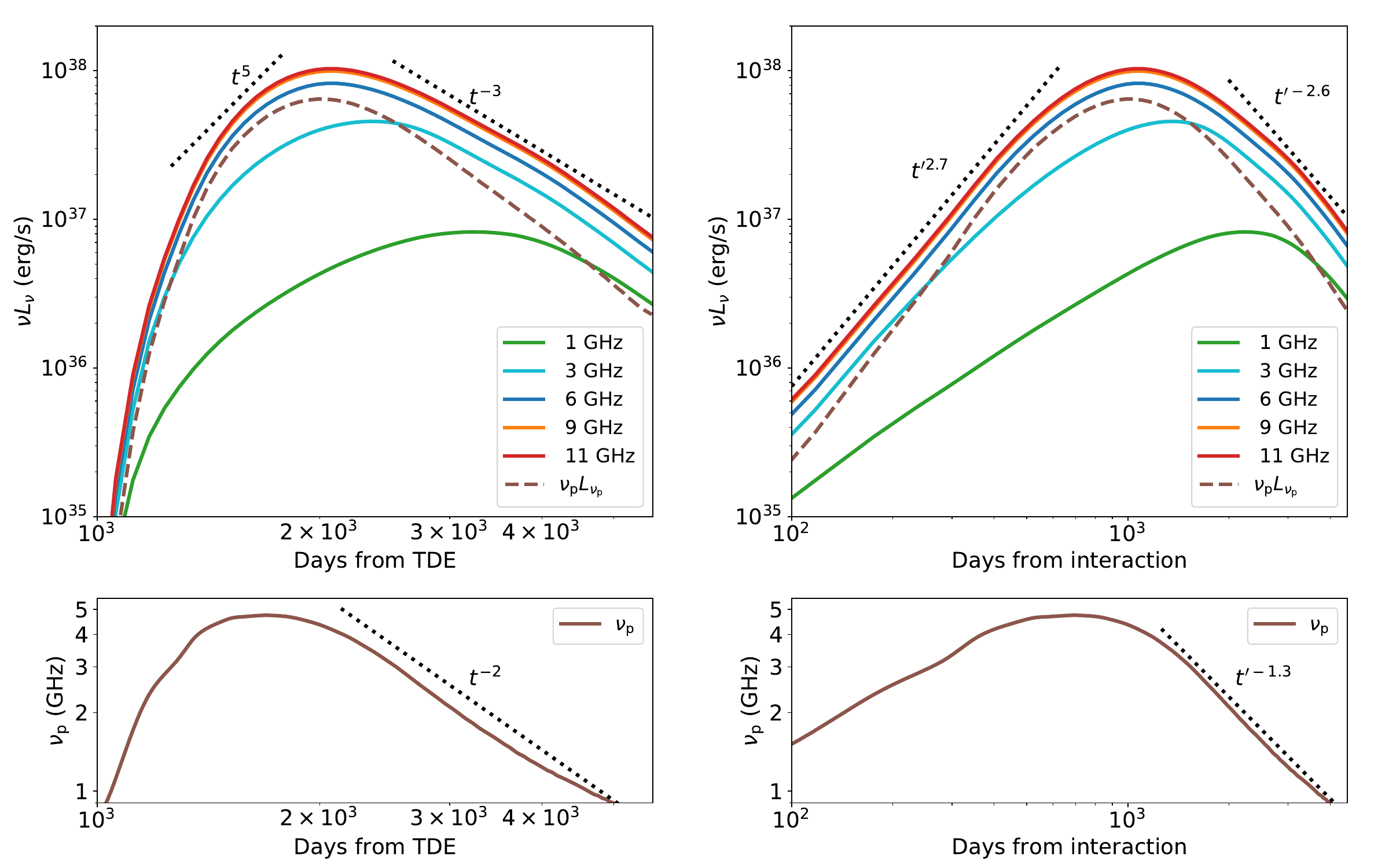}
  \caption{The top-left panel shows the radio light curves at different frequencies (solid lines), where the dashed line represents the light curve of $\nu_{\rm p}L_{\nu_{\rm p}}$. The evolution of the peak frequencies $\nu_{\rm p}$ is shown in the bottom-left panel. The right panel shows the same results but with the starting time set to the beginning of the interaction.}
  \label{fig:lc}
\end{figure*}

We calculate the radio emission from the shocked material according to Section \ref{sec:model} based on our simulation, where a face-on observer is assumed and it doesn't affect our results unless the shocked region is obscured by the torus. The shocked material will accelerate a fraction of non-thermal electrons, where we assume $\epsilon_{\text{e}} = 0.2$ and $\epsilon_\text{B} = 0.1$ in strong shocked material while $\epsilon_{\text{e}} = 0.02$ and $\epsilon_\text{B} = 0.01$ are adopted in weak shocked material due to the mixture with the internal torus material. As an example, we present the spectrum of the synchrotron radiation at different epochs in Figure \ref{fig:spc}, where $t=$1200 days roughly corresponds to the start time of the observable radio emission produced by the stream-torus interaction. The radio light curves at different frequencies are provided in the top-left panel of Figure \ref{fig:lc}, which normally follow a fast rise and a slow decay. In the initial rise stage before the peak luminosity, the radio light curve is much steeper than $\nu L_\nu \propto t^{5}$ (i.e., before $\sim$ 2000 days after the TDE event), when the shocked material starts to blow out after unbound debris collides with the torus. For the radio light curve, the high-frequency synchrotron emission rises faster, where peak luminosity occurs between 2110 and 3441 days from 11 GHz to 1 GHz. The peak time of the radio light curve at 1 GHz and 3 GHz in the top-left panel of Figure \ref{fig:lc} is different from those at other high-frequency wavebands, which is mainly caused by the optical-depth effect. For frequencies $\nu < 5\ \text{GHz}$, the peak time of the radio light curve corresponds to the transition of the shocked region from optically thick to optically thin. For frequencies $\nu > 5\ \text{GHz}$, peak time in the light curves is roughly simultaneous because the entire shock region is always optically thin.  
The peak radio luminosity at several GHz is $\sim 10^{38}$ erg/s for $M_{\rm BH}=10^{6}M_{\odot}$, where the light curve roughly follows $\nu L_\nu \propto t^{-3}$ after the peak, and becomes steeper over time, reaching $\nu L_\nu \propto t^{-3.5}$ (see top-left panel of Figure \ref{fig:lc}). The peak frequency in the synchrotron spectrum , $\nu_{\text{p}}$, also increases in the initial stage and roughly follows $\propto t^{-2}$ decay later on (see bottom-left panel). We present the contour lines corresponding to optical depths $\tau=1$ for different frequencies in the left panels of Figure \ref{fig:simulation}. During the declining phase, the deceleration during the dissipation process reduces the radiation intensity, and the spectral peak luminosity $\nu_{\text{p}} L_{\nu_{\text{p}}}$ declines with a rather steep slope, near $\propto t^{-4.0}$.



The main features of the light curve reflect the different stages of the interaction between the debris and the torus. After the collision between the debris and torus, the strong shocked material is produced, which will lead to a rapid increase of radio emission due to more shocked material. After the debris enters the cavity, the increase of strong shocked material will slow down. Furthermore, with the expansion of the shocked material, the magnetic field strength will become weaker and the fraction of nonthermal electrons become smaller due to the weaker shock, which will lead to the decay of the radio emission.


As follows, we present a simple analysis for the typical feature of the radio light curve in the decay phase, where we only consider the strong shocked material that contributes more than $90\%$ of the radio emission. 
A slight adjustment to the entropy criterion may slightly affect the timing and luminosity of the radio peak, but it will not impact the evolution of the decline phase of the light curve.
After the radio peak, the amount of strong shocked material will not increase significantly, and the expansion velocity of the shocked material evolves over time as $ v_{\text{ds}} \propto t'^{-k} $, where $t'$ is the time from the collision between the debris and torus. At this time, the dispersed shocked material can be simply approximated as a spherical shell expansion, where the size of the shocked material region is $ R\propto v_{\text{ds}} t' \propto t'^{1-k} $. The density of the shocked material then evolves as $ n_{\text{ds}} \propto R^{-2} \propto t'^{2k-2} $ if the thickness of the shell is not strongly varied. The density of nonthermal electrons is a fraction, $ (v_{\text{ds}}/v_{\text{DN}})^2 $, of total electron density, and, therefore, $ n_{\text{nt}} \propto n_{\text{ds}} v_{\text{ds}}^2 \propto t'^{-2} $ \citep{Sironi2013}. The magnetic field in shocked material follows $ B \propto (n_{\text{ds}} v_{\text{ds}}^3)^{1/2} \propto t'^{-(2+k)/2} $ (see Equation \ref{eq:B}). In the late stages, the light curves are entirely optically thin, so according to Equations \ref{eq:nu_m}, \ref{eq:L_nu_m}, and \ref{eq:L_nu}, we have $ L_\nu \propto v_{\text{ds}}^2 B^{(p+1)/2} \propto t'^{-(k+2)(p+1)/4-2k} $ for $ \nu > \nu_{\rm a} $. Similarly, the SSA frequency $ \nu_{\rm a} \propto (B^{(p+6)/2}/v_{\text{ds}})^{2/(p+4)} \propto t'^{-(k+2)(p+6)/2(p+4)+2k/(p+4)} $  based on Equation \ref{eq:nu_a}. After $\sim$1500 days from interaction, the total mass of the background material swept up by the strong shocked material becomes comparable to its own mass, and the velocity evolution near $ v_{\text{ST}} \propto t'^{-2/5} $ \citep[e.g.,][]{Taylor1950,Sedov1959}, which lead to $ L_\nu \propto t'^{-3(p+1)/5-4/5} $ and $ \nu_{\rm a} \propto t'^{-6(p+6)/5(p+4)+4/5(p+4)} $. For the typical value of $ p = 2.1 $, $ L_\nu \propto t'^{-2.7} $ and $ \nu_{\rm a} \propto t'^{-1.5} $, which is roughly consistent with that calculated from simulations (see right panels of Figure \ref{fig:lc}). It should be noted that the power-law index of the decay in the light curve is also related to the zero point of the starting time. The light curve in the radio band is normally plotted from the time of the optical/X-ray TDE event, which will lead to a steeper power-law index. As shown in the left panel of Figure \ref{fig:lc}, $L_\nu$ steeper than $ \propto t^{-3} $ and $\nu_{\text{p}}$ follows roughly a $\propto t^{-2}$ if the timing is measured from the starting time of optical/X-ray TDE. It is also worth noting that we simplified the debris to a straight cylinder rather than its actual parabolic shape, which means that the actual interaction area is approximately twice that of our current simulation. This would result in the production of more strongly shocked material, leading to higher luminosity and a slightly steeper rise, which will not affect the shape of the light curve.

\begin{figure*} 
  \centering
  \includegraphics[width=1.0\linewidth]{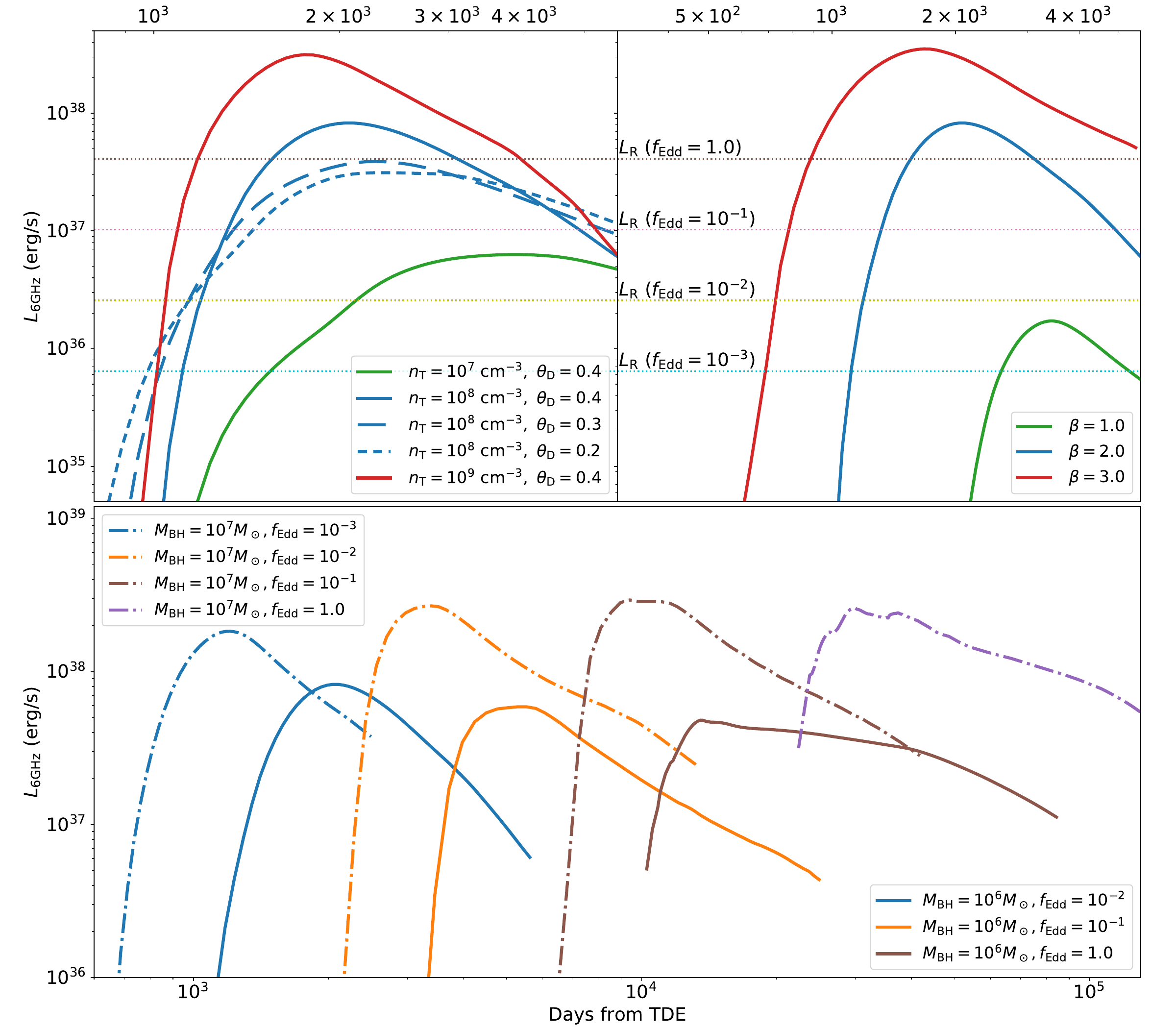}
  \caption{The top panel presents the radio light curves at 6 GHz for different model parameters, where the dotted lines present the typical radio emission from the radio quiet AGNs at different bolometric Eddington ratios with an SMBH mass of $M_{\rm BH}=10^{6}M_{\odot}$. The top-left panel presents the different torus densities and the injection angle for the debris stream with respect to the equatorial plane; The top-right panel shows the different penetration factors of TDEs; The bottom panel shows the results for the different SMBH masses and different Eddington ratios, which regulate the size of the dusty torus. The same color to represent the same dusty torus size.}
  \label{fig:parameterspace}
\end{figure*}

\begin{figure*} 
  \centering
  \includegraphics[width=1.0\linewidth]{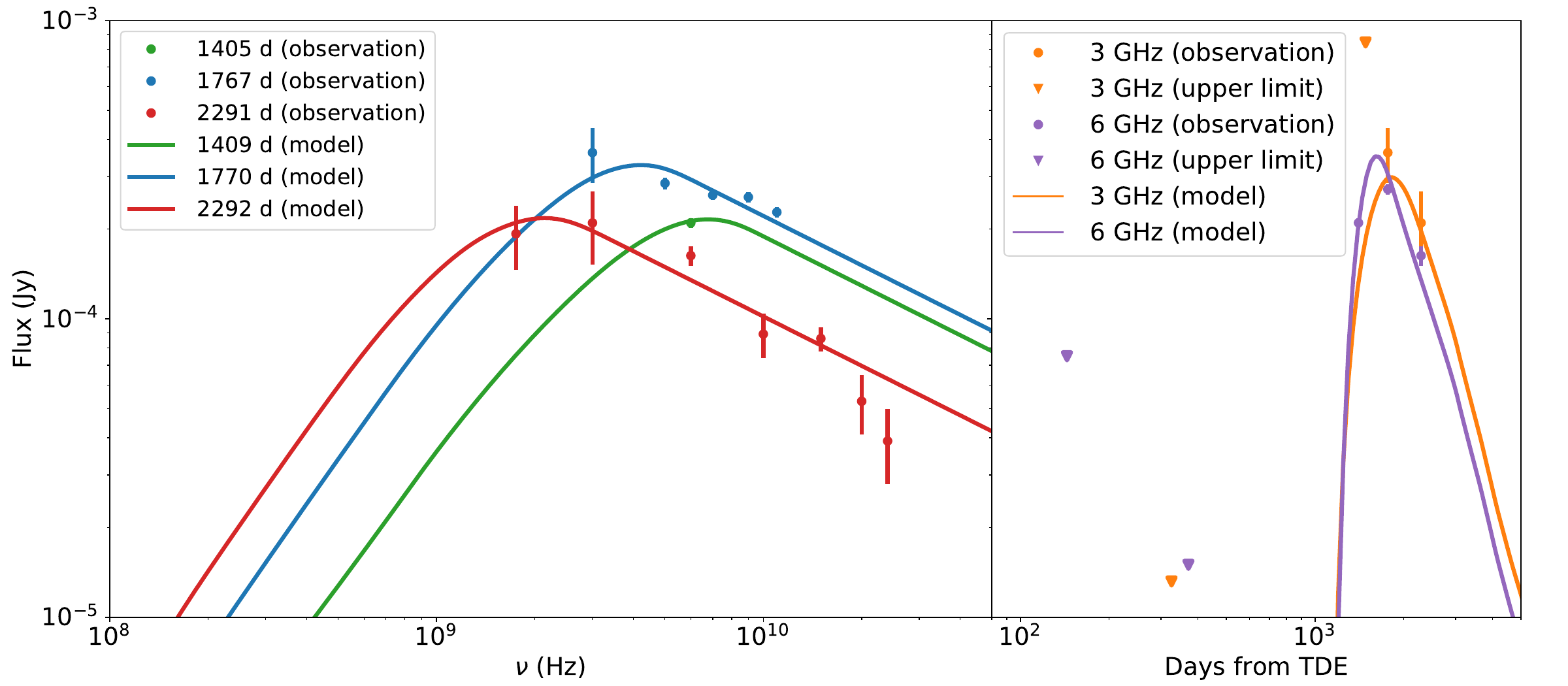}
  \caption{The left panel shows the spectral fits for PS16dtm on days 1767 and 2291 after detection. The right panel shows the radio light curves at 3 GHz and 6 GHz \citep[the data is selected from][]{Cendes2023}, where the solid lines are model predictions.}
  \label{fig:fit}
\end{figure*}

We find that the light curves from the lower resolution of $192 \times 256 \times 1024$ are roughly similar to that of $384 \times 512 \times 2048$ in our simulations. To explore the results with wider model parameter space, we present the results with a lower resolution of $192 \times 256 \times 1024$ in Figure \ref{fig:parameterspace}, where the model parameters are presented in Table \ref{table:result}. 

\begin{table}[h]
\centering
\caption{Model parameter space and part results}
\begin{tabular}{ccccc|cc}
\specialrule{0.3mm}{0pt}{0pt}
$M_{\text{BH}}$ & $f_{\text{Edd}}$ & $\beta$ & $n_\text{T}$ & $\theta_\text{D}$ & $t_{\text{peak}}$ & $(L_\text{6GHz})_{\text{peak}}$ \\ 
($M_\odot$) &   &   & ($\text{cm}^{-3}$) & (rad) & (days) & (erg/s) \\ \specialrule{0.3mm}{0pt}{0pt}
6        & 0.01             & 2       & $10^8$  & 0.4        & 2110              & $8.3 \times 10^{37}$  \\ \hline
6        & 0.01             & 2       & $10^8$  & 0.3        & 2275              & $3.9 \times 10^{37}$  \\ 
6        & 0.01             & 2       & $10^8$  & 0.2        & 2333              & $3.1 \times 10^{37}$  \\ \hline
6        & 0.01             & 2       & $10^9$  & 0.2        & 1755              & $3.1 \times 10^{38}$  \\ 
6        & 0.01             & 2       & $10^7$  & 0.2        & 3837              & $6.3 \times 10^{36}$  \\ \hline
6        & 0.01             & 3       & $10^8$  & 0.2        & 1675              & $3.5 \times 10^{38}$  \\ 
6        & 0.01             & 1       & $10^8$  & 0.2        & 3400              & $1.7 \times 10^{36}$  \\ \hline
6        & 1.0              & 2       & $10^8$  & 0.2        & 13526             & $4.8 \times 10^{37}$  \\ 
6        & 0.1              & 2       & $10^8$  & 0.2        & 5287              & $5.9 \times 10^{37}$  \\ 
7        & 1.0              & 2       & $10^8$  & 0.2        & 29767             & $2.6 \times 10^{38}$  \\ 
7        & 0.1              & 2       & $10^8$  & 0.2        & 9409              & $2.9 \times 10^{38}$  \\ 
7        & 0.01             & 2       & $10^8$  & 0.2        & 3389              & $2.7 \times 10^{38}$  \\ 
7        & 0.001            & 2       & $10^8$  & 0.2        & 1207              & $1.8 \times 10^{38}$  \\ \hline
\end{tabular}
\label{table:result}
\end{table}


In the top-left panel of Figure \ref{fig:parameterspace}, we present the radio light curve at 6 GHz for the different torus densities and different collision angles. We find that the higher torus density will lead to higher radio luminosity and a steeper rise with earlier epochs at given other parameters (solid lines), which is caused by the stronger collision and forms more shocked material in the case of the higher torus density. For a given torus density, it is found that the radio luminosity is also correlated with the collision angle. The smaller collision angle of $\theta_{\text{D}}$ will collide with the dusty torus at a smaller radius with a smaller interaction area considering the adopted torus structure, which will lead to a flatter light curve with earlier epoch and a slightly lower radio peak luminosity (blue lines). The specific peak luminosity at 6 GHz from different input parameters and the time to reach this peak luminosity after the TDE, is provided in Table \ref{table:result}.

In TDEs, both the initial binding energy and the ejection velocity of the unbound debris are related to the penetration factor $\beta$, where they are proportional to $\beta^2$ and $\beta$ respectively. We explore the effect of the penetration factor $\beta$ in the top-right panel of Figure \ref{fig:parameterspace}, where the higher penetration factor $\beta$ will lead to the higher initial binding energy and higher velocity of the unbound debris. It can be found that the maximum radio luminosity (see Table \ref{table:result}) increases more than two orders of magnitude when the $\beta$ value increases from 1 to 3, where the larger value of $\beta$ will also result in earlier radio emission, since that the arrival timescale is reduced at higher debris velocities. For comparison, we also present several typical radio luminosities of radio quiet AGNs at given SMBH mass, which is estimated from the fundamental plane of $\log L_\text{R} = 0.60 \log L_\text{X} + 0.78 \log (M_{\text{BH}}/M_\odot) + 7.33$ \citep{Merloni2003} by assuming  $L_\text{X} = L_{\text{bol}}/28$ \citep[e.g.,][]{Yang2015,Ho2008,Brightman2017}. It can be found that the radio flare triggered by the interaction between the unbound debris and the dusty torus is higher than the radio emission from radio quiet AGNs with typical Eddington ratio $L_{\rm bol}/L_{\rm Edd}\sim 0.1$, which is more evident in low-luminosity AGNs (e.g., $L_{\rm bol}/L_{\rm Edd} \le 0.01$). In particular, the timescale of the radio flare may last several years for a higher penetration factor (e.g., $\beta=$2-3, see the top-right plot of Figure \ref{fig:parameterspace}). 

In the bottom panel of Figure \ref{fig:parameterspace}, we present the radio light curves for different BH masses and Eddington ratios. It can be found that the delayed time, the peak radio luminosity and the possible decay timescale are closely correlated with both BH masses and Eddington ratios refer to Table \ref{table:result}, where these two parameters mainly affect the initial energy of unbound debris and the inner radius of the torus (see Equation \ref{eq:Rin}). The size of the torus will evidently affect the delay timescale of the radio emission, which is mainly regulated by the arriving time for the unbound debris to the torus (the dash-dot lines or solid lines), where the delay time for radio emission will be longer for the larger torus size. For a given penetration factor $\beta$, the higher BH mass will lead to a smaller disruption radius (in units of the gravitational radius) and higher kinetic energy for unbound debris. The higher-speed unbound debris will result in more shocked material or stronger radio emission, even when the inner radius of the torus remains the same (the lines with the same color in the bottom plot of Figure \ref{fig:parameterspace}). We find the duration of the radio outburst with Eddington ratios of $f_{\rm Edd} = 1.0$ is much longer than the other several cases with lower Eddington ratios. For example, the radio light curve roughly follows $\nu L_\nu \propto t^{-3.5}$ for $M_{\rm BH}=10^6 M_\odot$ with $f_{\text{Edd}} = 0.01$ ($R_{\text{in}} = 0.01$ pc), which become $\nu L_\nu \propto t^{-2}$ and $\nu L_\nu \propto t^{-0.5}$ for $f_{\text{Edd}} = 0.1$ and $f_{\text{Edd}} = 1.0$ respectively. The main reason is that the density of the unbound debris will become lower than the torus at the larger radius in the case of higher accretion rates, where the weaker collisions between debris and torus will lead to more shocked material flowing out. To shed light on this issue, we further take two extreme cases as an example. In the case of $M_{\rm{BH}} = 10^6 M_\odot$ and $f_{\rm{Edd}}=0.01$, the debris has a relatively high density when it collides with the torus, where $n_\text{d} \sim 10^{10} \ \text{cm}^{-3} \gg n_\text{T} \sim 10^8 \ \text{cm}^{-3}$. The debris can easily penetrate into the torus, with most of the subsequent interactions will occur within the torus. Most of the outflowing material is weak shocked material, which leads to a steeper decay. On the other hand, the debris has a lower density, $n_\text{d} \sim 10^6 \ \text{cm}^{-3} \ll n_\text{T} \sim 10^8 \ \text{cm}^{-3}$ in the case of $M_{\rm{BH}} = 10^7 M_\odot$ and $f_{\rm{Edd}}=1.0$, due to the longer propagation distance. In this case, most of the debris material is directly reflected upon impacting the torus surface, without significant interaction with the torus interior. As a result, most of the outflowing gas is strong shocked material, which leads to a slow and prolonged decay.


\subsection{Application to PS16dtm} \label{sec:PS16dtm}

We apply our model to a TDE candidate of PS16dtm \citep{Chambers2016}, which is found in a narrow-line Seyfert 1 (NLSy1) galaxy at z = 0.08 on August 12, 2016. The central SMBH mass is $\sim 10^6 M_\odot$ \citep{Greene2007, Xiao2011, Blanchard2017}. In the last several years, it has accumulated a substantial amount of multi-wavelength observational data \citep{Blanchard2017, Jiang2017, Petrushevska2023, Cendes2023}. \citet{Cendes2023} reported the first radio signal at 6 GHz, reaching $2 \times 10^{38}$ erg/s at 1405 days after the reported TDEs, followed by additional radio signals at 1767 days and 2291 days at different frequencies. \citet{Blanchard2017} proposed that the disrupted stellar mass $M_\text{s} \sim 0.22 M_\odot$ based on the fitting of the X-ray light curve. Based on the echo of infrared emission, \citet{Jiang2017} constrained the dusty torus size $\sim 9 \times 10^{-3}$ pc. These observational constraints will be adopted in our simulations.

In Figure \ref{fig:fit}, we present the fitting of both the radio spectra and radio light curves for PS16dtm, where $n_{\rm T} = 5 \times 10^8 \text{cm}^{-3}$, $\epsilon_{\text{e}} = 0.1$, $\epsilon_\text{B} = 0.01$ in strong shocked material and  $\epsilon_{\text{e}} = 0.05$, $\epsilon_\text{B} = 0.005$ in weak shocked material respectively. In the left panel, it can be found that the spectra at three epochs can be well reproduced. In the right panel, the model predicts the fast rise of the radio light curves, which is also consistent with the observations of PST16dtm. Our model can roughly reproduce the radio spectra at different days (see left panel of Figure \ref{fig:fit}). The right panel of Figure \ref{fig:fit} presents the radio light curves of PST16dtm, where the model prediction is consistent with the observations very well.

\section{Conclusion and Discussion} \label{sec:conclusion&discussion}

When a star gets too close to the SMBH, it will be torn apart by the tidal forces. Roughly half of the stellar mass becomes unbound and flies away at tremendous velocities of $\sim 10^4\rm km/s$. The debris stream possibly collides with the geometrically thick dusty torus in the nuclear region of galaxies (especially in AGNs), which can trigger a transient flare. Based on the three-dimensional hydrodynamic simulations, we explore the synchrotron radiation from the shocked material from the interaction of the unbound debris with the surrounding dusty torus. The main results include: 1) Due to the continuous collision between the unbound debris and the dense dusty torus, a fraction of the shocked material will be spilled out from the interacting gap and form an outflow. The non-thermal electrons can be formed in the shocked material, which will contribute to synchrotron radiation; 2) The radio emission from the synchrotron radiation will delay the optical/X-ray outburst for about several years or even several tens years, where the delay timescale is determined by the velocity of the debris and the size of dusty torus; 3) The radio light curve in this model normally follows a steep rise and slow decline due to the high density of the cold dusty torus, which is much different from the prediction of the collision between the outflow/jet with the CNM; 4) This model can reproduce both the radio spectrum and the light curve well in a narrow-line Seyfert I of PS16dtm. 


The origin of the multi-waveband radiation of TDEs is still in hot debate now. The radio emission is normally associated with synchrotron emission that either from the relativistic jets (e.g., Sw J1644+57\citep[e.g.,][]{Bloom2011,Zauderer2011,Levan2011,Metzger2012}, Sw J2058+05 \citep[e.g.,][]{Cenko2012}, Sw J1112-82 \citep[e.g.,][]{Brown2015} and AT 2022cmc \citep[e.g.,][]{Andreoni2022}.) or the interaction of the outflowing material with the surrounding medium \citetext{\citealp[e.g.,][]{Alexander2016,Krolik2016}; \citealp[see][for review]{Alexander2020}}. Several channels can launch non-relativistic outflows from TDEs and each of them can potentially produce the observed radio emission. The first channel is disk winds that formed from the super-Eddington accretion disk \citep[e.g.,][]{Strubbe2009,Metzger2016,Cao2022}. The second possible channel is from the debris stream collision \citep[e.g.,][]{Jiangyf2016,Lu2020}. The third channel is the outflowing unbound debris, which is also launched at the moment of disruption \citep{Krolik2016,Yalinewich2019}. The radiation from the interaction of the outflow with CNM surrounding the SMBHs has been explored in former works \citetext{\citealp[e.g.,][]{Alexander2016,Krolik2016}; \citealp[see][for review]{Alexander2020}}. The interaction between wind/debris with the CNM produces a shock front where the magnetic field is amplified and electrons are accelerated to relativistic energy. In this work, we explore another channel where we consider the outflow triggered by the collision between the debris stream and dusty torus, where the density of cold torus is normally much higher than that of CNM. 
Therefore, our model predicts a very steep radio light curve compared to that in the interaction of outflow and CNM (see Figures \ref{fig:lc}, \ref{fig:parameterspace}). Currently, some TDEs have been observed with unique late-time fast-rise radio light curves. For example, AT2018hyz was initially detected in radio wavebands approximately 1000 days after its optical discovery, which shows a very steep rise with $L_\nu \propto t^{5}$ \citep{Cendes2022}. The host galaxy of AT2018hyz was proposed an a weak AGN with$ \sim 0.1 - 1 \% $ Eddington luminosity for the inferred black hole mass of $\sim 10^{6-7} M_\odot$ \citep[e.g.,][]{Aird2015,Gomez2020,Short2020}. ASASSN-15oi also show a very steep radio rise with $L_\nu \propto t^{4}$ and a subsequent steep decline $L_\nu \propto t^{-3}$ \citep{Horesh2021a}. The early observations of ASASSN-15oi also do not rule out the possibility of a low-luminosity AGN \citep[e.g.,][]{Holoien2016b,Gomez2020}. More observations on the host galaxies of TDEs will shed light on the possible origin of radio emission.
Furthermore, the time delay of our model is normally much longer than those of other models due to the not very high escape velocity of debris and the large size of cold torus, which provide a possibility to explain the long delayed radio outbursts in some TDEs \citep[e.g.,][]{Alexander2016,Cendes2021,Cendes2021a,Horesh2021a,Stein2021,Matsumoto2022,Matsumoto2023,Beniamini2023}.


Several other models were also proposed to explain the fast rise and the long delay of radio emission, such as the outflow/jet-ISM interaction or the off-axis jet model. Both one-component and two-component off-axis jet models can produce fast-rising, slowly declining late-time radio light curves, with extremely steep rises that can even exceed $L_\nu \propto t^{5}$, effectively explaining sources like AT 2018hyz \citep[e.g.,][]{Matsumoto2023,Sfaradi2024,Sato2024}. These jet models normally predict shallower radio light curve in the decay phase \citep[e.g., usually near $L\propto t^{-2}$ or more shallower][]{Cendes2021,Horesh2021a,Matsumoto2023}, which may be also parameter dependent. The interaction of low-velocity wind/outflow and CNM is normally mild, which leads to slow rise and slow decay in light curve \citep[e.g., near $L_\nu \propto t^{2.5}$ and $L_\nu \propto t^{-1.5}$,][]{Krolik2016,Matsumoto2021,Matsumoto2024}. The interaction between the outflow and the clumpy torus or gaseous clouds normally predicts a much shallower rise and decline in the light curve \citep[e.g.,][]{Mou2022,Bu2023}. \citet{Zhuang2024} proposed that the interaction between outflows and single dense cloud can generate late-time fast-rise radio flares. Similar to our model, this interaction leads to rapid shock formation due to collisions, causing a steep rise, and a rapid decline due to cooling and expansion, where the dense cloud outside is more or less similar to the torus of our model. The simultaneous modeling for the spectrum and rise/decay radio light curve will be helpful to constrain these models. Furthermore, high-resolution VLBI observations can help to distinguish the relativistic jet or the low-velocity wind/outflow model.

We consider the collision of debris streams with the dusty torus in the nuclear region of galaxies. There is no doubt that the dusty torus exists in AGNs, which is a key gradient in AGN unification model \citep[e.g.,][]{Antonucci1993,Urry1995,Netzer2015}. The gas/dust will move inward as the BH accretion rate decreases, where the torus size becomes smaller in low-luminosity AGNs \citep[e.g.,][]{Pier1992,Hatziminaoglou2008,Kawaguchi2011,Kishimoto2011}. It is still unclear how much cold gas and dust is in the nuclear region of normal galaxies. Several recent studies show that TDE hosts are significantly overrepresented in post-starburst or E+A galaxies, whose spectra are characterized by a lack of strong emission lines but with strong Balmer absorption features \citetext{\citealp[e.g.,][]{Yao2023,Wang2024}; \citealp[see][for review]{French2020}}. It is believed that the post-starburst galaxies may belong to a fading stage of former AGNs \citep{Goncalves1999,Goto2006,Baron2017,French2023}. If this is the case, the cold gas/dust may still exist in the center of galaxies, which should be difficult to observe due to the turn-off of the SMBH activities. There is some evidence for the presence of strong dust in some TDEs that are found in quiescent galaxies, where the strong infrared radiations are detected after the optical TDEs \citep[e.g.,][]{Stein2021,van_velzen2021b, Masterson2024,Hinkle2024}. This implies that similar processes, such as interactions between debris and a dusty torus, could also occur in these galaxies. Our model may provide an explanation for the second radio burst as found a thousand days later after the optical TDE of AT 2019dsg with strong infrared radiation \citep{van_velzen2021b,Jiang2023,Cendes2023}. Therefore, apart from the TDEs in AGNs or low-luminosity AGNs, the collision between the debris stream and dusty torus is still possible in TDEs that happened in normal galaxies. 

More and more detections of the late radio emission are reported in TDEs. Considering the possibility of the interaction between unbound debris and the torus, we employed numerical simulations to model this interaction process. The hyperbolic orbits of the unbound debris will sweep a larger area compared to that of the assumed straight cylindrical flow during the collision of the debris and dusty torus, which will lead to more shocked material and stronger radio emission. The wide-field survey of next-generation radio telescopes such as SKA, will detect more radio flares of TDEs, which will further help us to distinguish the different channels for the radio outbursts and explore the galactic nuclear environment.

\begin{acknowledgments}
We appreciate the referee for very constructive suggestions. We are grateful to Guobin Mou and Di Wang for their helpful discussions. The work is supported by the National SKA Program of China (No. 2022SKA0120101), National Natural Science Foundation of China (grants 12233007 \& grants 12473012), and the National Key Research and Development Program of China (No. 2023YFC2206702). H.L. gratefully acknowledges the support by  LANL/LDRD under project number 20220087DR. Y.P.L. is supported in part by the Natural Science Foundation of China (grants 12373070, and 12192223), the Natural Science Foundation of Shanghai (grant NO. 23ZR1473700). The authors acknowledge Beijing PARATERA Tech CO., Ltd. for providing HPC resources that have contributed to the results reported within this paper.
\end{acknowledgments}

\bibliography{cite}{}

\bibliographystyle{aasjournal}

\end{document}